\documentclass[conference]{IEEEtran}
\IEEEoverridecommandlockouts
\usepackage{cite}
\usepackage{amsmath,amssymb,amsfonts}
\usepackage{algorithmic}
\usepackage{algorithm}
\usepackage{graphicx}
\usepackage{epstopdf}
\usepackage{textcomp}
\usepackage{xcolor}
\usepackage{bm}
\usepackage{booktabs}
\usepackage{xcolor}
\usepackage{colortbl, booktabs}
\usepackage{multirow}
\usepackage{subcaption}
\usepackage{float}
\usepackage{makecell}

\def\BibTeX{{\rm B\kern-.05em{\sc i\kern-.025em b}\kern-.08em
    T\kern-.1667em\lower.7ex\hbox{E}\kern-.125emX}}

\begin{document}

\title{Data Fusion for BS-UE Cooperative \\MIMO-OFDM ISAC}

\author{
    \IEEEauthorblockN{
        Yixin Ding\textsuperscript{1}, 
        Haoyu Jiang\textsuperscript{1}, 
        Xiaoli Xu\textsuperscript{1}, 
        Yanan Liang\textsuperscript{2}, 
        Yong Zeng\textsuperscript{1,3}
    }
     \IEEEauthorblockA{\textsuperscript{1}National Mobile Communications Research Laboratory, Southeast University, Nanjing, China\\
    \IEEEauthorblockA{\textsuperscript{2}School of Electronic Information Engineering, Beijing Jiaotong University, Beijing, China}
    \IEEEauthorblockA{\textsuperscript{3}Purple Mountain Laboratories, Nanjing, China}
    Emails: 13606141969@163.com,
    ynliang@bjtu.edu.cn, \{213211563; xiaolixu; yong\_zeng\}@seu.edu.cn}
}

\maketitle
\begin{abstract}
Integrated sensing and communication (ISAC) is a promising technique for expanding the functionalities of wireless networks with enhanced spectral efficiency. The 3rd Generation Partnership Project (3GPP) has defined six basic sensing operation modes in wireless networks. To further enhance the sensing capability of wireless networks, this paper proposes a new sensing operation mode, i.e., the base station (BS) and user equipment (UE) cooperative sensing. Specifically, after decoding the communication data, the UE further processes the received signal to extract the target sensing information. We propose an efficient algorithm for fusing the sensing results obtained by the BS and UE, by exploiting the geometric relationship among BS, UE and targets as well as the expected sensing quality in the BS monostatic and BS-UE bistatic sensing. The results show that the proposed data fusion method for cooperative sensing can effectively improve the position and velocity estimation accuracy of multiple targets, and provide a new approach on the expansion of the sensing pattern.
\end{abstract}


\section{Introduction}
Integrated sensing and communication (ISAC) aims to integrate the sensing functionality into the wireless communication system, enabling simultaneously transmitting information and conducting environmental sensing. This approach achieves more efficient use of limited spectrum resources and reduces the need for separate sensing equipment. The International Telecommunication Union Radiocommunication Sector (ITU-R) officially listed ISAC as one of the six major application scenarios for 6G networks \cite{ITU}. In addition, 3GPP has discussed six deployment scenarios for ISAC \cite{3GPP_ISAC}: (1) BS monostatic (2) BS-BS bi-static (3) UE monostatic (4) UE-UE bi-static (5) BS-UE bi-static and (6) UE-BS bi-static. Broadly speaking, they can be divided into two categories: monostatic and bi-static. 

Monostatic sensing has the advantage of easy synchronization because the sensing transmitter and receiver are collocated. But its range is limited and needs sufficient antenna separation to avoid self-interference. Bi-static sensing, in contrast, can effectively expands the sensing range and obtain more target information through the cooperation of two nodes\cite{Brunner2024BistaticOI}. However, with transmitter and receiver separated, clock synchronization and symbol transmission accuracy are hard to ensure. Thus, several studies focus on solving the problems of these individual sensing modes \cite{Liu2021IntegratedSA}.

On the other side, given that each type of sensing mode has its own unique strengths and weaknesses, exploring the integration of multiple sensing modes has emerged as a novel perspective for addressing such issues. Researches indicate that the integrated approach can not only reduce the burden on individual sensing methods, but also eliminate existing problems through proper data fusion algorithms \cite{Xie2022CollaborativeSI}. Reference \cite{Jiang2023CooperationBasedJA} investigated combing BS monostatic sensing and BS-BS bi-static sensing. It showed that correlating active and passive sensing information could reduce the estimation errors of time offset (TO) and carrier frequency offset (CFO). Additionally, it utilized the fractional Fourier transform (FRFT) algorithm to improve positioning accuracy. For data fusion strategy, the authors in \cite{Zhou2024JointTD} proposed a novel sparse Bayesian learning framework to extract sparsity of both common and individual information. Through distributed processing and a fast hybrid message passing algorithm, the transmission efficiency has been improved. \cite{9727176} focused on the perceptive mobile network for UE, which realized mutual enhancement between active and passive perception by extending the belief propagation (BP) method based on hybrid simultaneous localization and mapping (SLAM). 

While cooperative sensing holds promise for enhancing sensing performance, its practical implementation faces numerous challenges, including spatial registration, time synchronization, and communication overhead \cite{10570616}. In this paper, we propose an effective cooperative framework between the BS and UE, by fusing the sensing results obtained from the BS monostatic and BS-UE bi-static sensing. To relax the synchronization requirement and reduce the communication overhead, we assume that the UE will only send the sensing results to BS for cooperative sensing. We utilize the BS sensing results to obtain the initial target position estimation. Due to the limited resolution of the Fast Fourier Transform (FFT) and the variability of channels, the observed values from the BS and UE usually deviate from the ground-truth values. To minimize the sensing error via cooperation, all observed values of the BS and UE are linked to the real geometric relationship between BS, a target and UE to form the expressions of errors. Solving for optimized target values then reduces to addressing a Least Squares (LS) problem. We establish a target function based on the error vector and the weighted matrix, through which target position and velocity estimates are optimized via iteration. Simulation results show that incorporating UE sensing results improves target positioning accuracy to the decimeter level. Additionally, this approach enables the estimation of both radial and tangential velocities, rather than merely the radial velocity measurable by the BS.
\begin{figure}[t]
\centerline{\includegraphics[width=0.9\columnwidth]{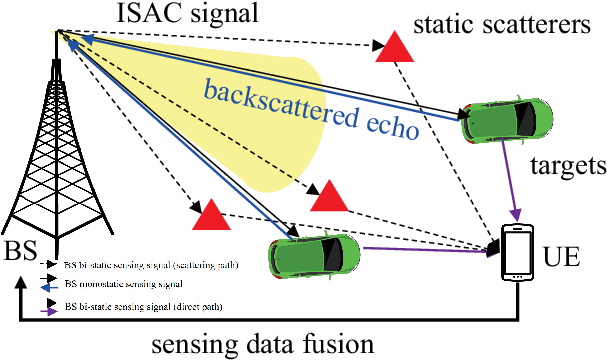}}
\caption{BS-UE cooperative sensing for ISAC.}
\vspace{-\baselineskip}
\label{F:model}
\end{figure}

\section{System Model}
As shown in Fig.~\ref{F:model}, we consider a communication-centric ISAC system, where the BS aims to sense the environment while communicating with the UE. Among the six sensing modes defined by 3GPP \cite{3GPP_ISAC}, two are involved in this scenario: BS monostatic sensing and BS-UE bi-static sensing.

In general, the BS has higher sensing capability due to the larger antennas equipped and the prior information on the embedded data. However, the backscattered signal received by the BS is usually weak due to the propagation of the dual hop wave and the long travel distance. While the UE can receive stronger signals from nearby sensing targets, but its spatial resolution is low due to the much smaller antenna size. Besides, the data decoding error will also degrade the sensing accuracy. To this end, we propose to enhance the sensing performance by fusing the sensing results of the BS and UE. 

\subsection{Channel Models}
Without loss of generality, we assume that the BS locates at the origin and the location of the UE is denoted by $\mathbf{q}_{U}=(x_{U},y_{U})^T\in\mathbb{R}^{2\times 1}$. There are $L_t$ moving targets within the area of interest, where the $l$th target locates at $\mathbf{q}_l=(x_l,y_l)^T\in\mathbb{R}^{2\times 1}$, and it moves with velocity $\mathbf{v}_l=(v_{x,l},v_{y,l})^T\in\mathbb{R}^{2\times 1}$, $l=1,2,...,L_t$. Besides the targets, there are $L_p$ randomly located static scatterers with position $\mathbf{q}_s$. 

Denote by $M_T$ and $M_R$ the number of transmit and receive antennas at the BS, respectively. The impulse response for the backscattered channel can be represented by  $\bm{H}_{BB}(t,\tau)\in\mathbb{C}^{M_R\times M_T}$, which is expressed as
\begin{align}
\bm{H}_{BB}(t,\tau)&=\sum_{l=1}^{L_t}\alpha_{l}\bm{\beta}_{r}(\mathbf{q}_l)\bm{\beta}_{t}^T(\mathbf{q}_l)\delta(\tau-\tau_{B,l})e^{j2\pi f_{D,l}^Bt}\nonumber\\
&+\sum_{s=1}^{L_p}\alpha_s\bm{\beta}_{r}(\mathbf{q}_s)\bm{\beta}_{t}^T(\mathbf{q}_s)\delta(\tau-\tau_{B,s})\label{eq:HBB},
\end{align}
where $\alpha_l$ and $\alpha_s$ are the complex channel gain for the path reflected from the $l$th target and $s$th scatterer respectively. Channel gain amplitudes depend on travel distance and target/scatterer RCS, and phases are usually assumed random.
$\bm{\beta}_t(\mathbf{q})\in\mathbb{C}^{M_T\times 1}$ and $\bm{\beta}_r(\mathbf{q})\in\mathbb{C}^{M_R\times 1}$ are the BS transmit and receive steering vector towards/from the location $\mathbf{q}$. With uniform linear array (ULA), $\bm{\beta}_t(\mathbf{q}_l)$ can be given by:
\begin{equation}
\bm{\beta}_{t}(\mathbf{q}_l) = \begin{bmatrix}
1,e^{-j\frac{2\pi d}{\lambda}\sin\theta_{B,l} },...,e^{-j\frac{2\pi d(M_T-1)}{\lambda}\sin\theta_{B,l} }
\end{bmatrix}^T,\label{eq:BR}
\end{equation}where $d$ is the antenna array spacing, $\theta_{B,l}=\arctan(x_l/y_l)$ is the angle of arrival/departure (AoA/AoD) the $l$th target with BS. The expression of $\bm{\beta}_r(\mathbf{q})$ is similar, substitute $M_T$ with $M_R$. The signal propagation delays are given by $\tau_{B,l}=\frac{2 \| \mathbf{q}_l \|}{c_0},\tau_{B,s}=\frac{2 \| \mathbf{q}_s \|}{c_0},$ where $c_0$ is the speed of light. As the targets are moving, the corresponding paths also introduce a Doppler shift. Specifically, the Doppler shift for the $l$th target is $f_{D,l}^{B} = \frac{2 \|\mathbf{v}_l\|\cos\theta_l^B}{c_0} f_c,$, which is related with the target's radiate velocity. Here, $\theta_l^B$ denotes the angle between the transmitter's radar line of sight and the moving direction of the $l$th target, and $f_c$ is the carrier frequency. 

Next, we consider the channel between the BS to UE. A rich scattering environment is assumed and hence the line-of-sight link between the BS and UE is absent. For simplicity, we assume that the UE is equipped with a single antenna and hence the channel impulse response can be represented by $\bm{h}_{BU}(t,\tau)\in\mathbb{C}^{M_{T}\times 1}$, written as
\begin{align}
\bm{h}_{BU}(t,\tau)&=\sum_{l=1}^{L_t}\alpha_{l,u}\bm{\beta}_{t}(\mathbf{q}_l)\delta(\tau-\tau_{U,l})e^{j2\pi f_{D,l}^Ut}\nonumber\\
&+\sum_{s=1}^{L_p}\alpha_{s,u}\bm{\beta}_{t}(\mathbf{q}_s)\delta(\tau-\tau_{U,s})\label{eq:hBU},
\end{align}
where the channel gains $\alpha_{l,u}$, $\alpha_{s,u}$ and transmit steering vector are defined similarly as in \eqref{eq:HBB}. The path delays are given by $\tau_{U,l}= \frac{ \| \mathbf{q}_l \|+ \| \mathbf{q}_{U} - \mathbf{q}_l \|}{c_0},
\tau_{U,s}= \frac{ \| \mathbf{q}_s \|+ \| \mathbf{q}_{U} - \mathbf{q}_s \|}{c_0}.$
The Doppler shift introduced by the moving target is $f_{D,l}^{U} = \frac{2(\|\mathbf{v}_l\| \cos\theta_l^B+\mathbf{v}_l \cos\theta_l^U)}{c_0} f_c,$ where $\theta_l^U$ is the angle between the moving direction of the $l$th target and the receiver.

\subsection{Signal Model}
We assume that the ISAC signal shown in Fig.~\ref{F:model} is modulated with OFDM. Denote the OFDM subcarrier spacing by $\Delta f$, and the OFDM symbol duration by $T$. With $T=1/\Delta f$, the orthogonality between different subcarriers can be guaranteed. To avoid inter-Symbol interference (ISI), cyclic prefix (CP) with length $T_{cp}$ is added for each OFDM symbol. Hence, the effective OFDM symbol length is $T_s=T+T_{cp}$. The transmitted OFDM signal by $M_T$ antennas can be written by
\begin{align}
\small
\mathbf{x}(t) = \sum_{m = 0}^{M - 1}\sum_{k = 0}^{N - 1} \bm{w}_{k,m}d_{k,m} e^{j2\pi k\Delta f(t - mT_s - T_{cp})} \mathrm{rect}\left(\frac{t - mT_s}{T_s}\right),
\end{align}where $N$ is the  number of subcarriers, $M$ is the number of OFDM symbols, and $d_{k,m}$ is the data carried by the $k$th subcarrier of the $m$th symbol. $\bm{w}_{k,m}\in \mathbb{C}^{M_T\times 1}$ is the transmit beamforming vector.

For notational convenience, we represent the transmit signal $\mathbf{x}(t)$ in the space-frequency-time domain by the tensor $\mathcal{X}\in \mathbb{C}^{M_T\times N\times M}$, where $[\mathcal{X}]_{i,k,m}=\bm{w}_{k,m}(i)d_{k,m}$ denotes the equivalent symbol sent on the $i$th antenna element, $k$th subcarrier and $m$th OFDM symbol. 

Next, we consider the channel frequency-domain representation in \eqref{eq:HBB} and \eqref{eq:hBU}. Following the convention, we assume that the Doppler shift is approximately constant during each OFDM symbol duration. Remove CP and convert signals from serial to parallel for each OFDM symbol. Then, through $N$-point DFT on the subcarrier direction using cyclic property on channel response, we can obtain the channel matrices: $\tilde{\bm{H}}_{BB}^{km}\in \mathbb{C}^{M_R\times M_T}$ and similarly derive $\tilde{\bm{h}}_{BU}^{km}\in \mathbb{C}^{M_T\times 1}$ as 
\begin{equation}
\tilde{\bm{H}}_{BB}^{km}=FFT\{{\bm{H}}_{BB}(mTs,n\frac{N}{T})\}_n.
\end{equation}

The received signals at the BS and the UE after OFDM demodulation can be represented by $\mathcal{Y}_B\in\mathbb{C}^{M_R\times N\times M}$ and
$\mathbf{Y}_U\in\mathbb{C}^{N\times M}$, respectively. They are related with the transmitted signal and channel as
\begin{align}
\mathcal{Y}_B = \mathcal{X} \times_1 \tilde{\bm{H}}_{BB}^{km} + \mathcal{N}_{B},
\mathbf{Y}_U = \mathcal{X} \times_1 \left(\tilde{\bm{h}}_{BU}^{km}\right)^T + \mathbf{N}_{U}. \label{eq:combined}
\end{align}where $\times_1$ means tensor $\mathcal{X}$ is multiplied by the matrix $\tilde{\bm{H}}_{BB}^{km}$ and $\left(\tilde{\bm{h}}_{BU}^{km}\right)^T$in the first mode. $\mathcal{N}_{B}\in \mathbb{C}^{M_R\times N\times M}$ and $\mathbf{N}_{U}\in \mathbb{C}^{N\times M}$ are noise term observed at the BS and UE receivers. 

\section{BS-UE cooperative sensing}
We propose fusing BS and UE sensing results to enhance performance, since target info can be estimated from the BS-observed reflected channel and the UE-observed direct channel. For brevity, the noise term are neglected in the following. Moreover, to get accurate sensing results, interferences from stationary scatterers and BS itself in the received signal are mitigated using the approach described in \cite{b5}.

Upon receiving the reflected signal $\mathcal{Y}_B$, the BS first removes the communication data symbol to obtain the data cube for sensing. Specifically, the BS uses the transmitted data stored at the buffer to divide the received signal as
\begin{align}
 \left[\mathcal{F}_B\right]&_{:,k,m} = \frac{[\mathcal{Y}_B]_{:,k,m}}{d_{k,m}}\nonumber\\
&=\sum_{l=1}^{L_t}\tilde{\alpha}_{l,k,m}\cdot \bm{\beta}_r(\mathbf{q}_l) e^{-j2\pi k\frac{\tau_{B,l}}{T}}e^{j2\pi f_{D,l}^BmT_s},   
\end{align} where $[\mathcal{Y}_B]_{:,k,m}$ is the received signal on all the antenna elements, $k$th subcarrier and $m$th symbol, and $\tilde{\alpha}_{l,k,m}=\alpha_l\bm{\beta}_t^T(\mathbf{q}_l)\bm{w}_{k,m}$ is the effective complex channel gain for the $l$th target. Since the BS has multiple antennas, it has the ability to distinguish angles the targets. For high resolution, the Multiple Signal Classification (MUSIC) algorithm is employed with Minimum Description Length (MDL) method for signal number detection.

Then, the BS conducts beamforming towards the estimated angle $\hat{\theta}_{B,l}$ for each target. The beamforming vector $\bm{\beta}_r(\hat{\theta}_{B,l})$ is defined in a manner similar to~(\ref{eq:BR}). If the estimated angle is accurate, i.e, $\hat{\theta}_{B,l}={\theta}_{B,l}$, the beamforming gain of $M_R$ is obtained for the $l$th target and the gains for targets on other directions are usually smaller. The received matrix towards the $l$th target is represented by a matrix $\bm{F}_{B,l} \in\mathbb{C}^{N\times M}$, which is
\begin{equation}
\bm{F}_{B,l}={\mathcal F}_B\times_1\bm{\beta}_r(\hat{\theta}_{B,l}).
\end{equation}

Then, we use periodogram to estimate the time delay $\tau_{B,l}$ and Doppler $f^{B}_{D,l}$. By applying $N$-point IFFT along the OFDM subcarrier direction and an $M$-point FFT along the symbol direction to $\bm{F}_{B,l}$, the Range-Velocity mapping relationship can be obtained.

The resultant delay-Doppler profile $\mathcal{C}_{B,l}$ is a complex matrix of size $N\times M$ and the peak gives the estimated range and Doppler of the $l$th target. Denote the index of the highest peak by $(\hat{n}_l,\hat{m}_l)$, the delay and Doppler introduced by the $l$th target are estimated as $\hat{\tau}_{B,l} = \frac{\hat{n_l}}{N\Delta f}, \hat{f}_{D,l}^{B} = \frac{\hat{m_l}}{M T_s}$.

Repeating the above sensing process for all the $L_t$ targets, the BS can get the set of target parameters as
\begin{equation}
\Psi_B=\left\{ \left(\hat\tau_{B,l}, \hat{f}_{D,l}^{B}, \hat\theta_{B,l})\right), l = 1, \ldots, L_t \right\}. 
\end{equation}

\subsection{BS-UE Bi-static Sensing}
When the targets are close to the UE, the signal received by the UE maybe stronger than the echoes received by the BS. This allows the UE to correctly detect the targets and obtain better estimation results.  For communication purposes, the UE first decodes the data embedded in the signal. We assume that the UE can perfectly decode the data $d_{k,m}, \forall k,m$. For sensing purpose, the data will be removed from the received signal as the reference transmitted data, which renders the sensing information cube: 
\begin{align}
\mathbf{F}_{U_{k,m}} &= \frac{\mathbf{Y}_{U_{k,m}}}{d_{k,m}}
=\sum_{l=1}^{L_t}\tilde{\alpha}_{l,k,m}e^{-j2\pi k\frac{\tau_{U,l}}{T}}e^{j2\pi f_{D,l}^U m T_s}.  
\end{align}

Similar to signal processing in BS monostatic sensing, the interference from stationary scatterers is eliminated by the frequency domain filters, and then 2-D FFT is adopted to estimate the delay and Doppler of targets. The details are omitted for brevity and the estimated results at the UE are denoted by 
\begin{align}
\Psi_U=\{ (\hat\tau_{U,l}, \hat f_{D,l}^{U} ), l = 1, \ldots, L_t \}.\label{eq:EstU}
\end{align}

Then, for data fusion, the estimated UE results are sent to the BS via the uplink control channel. 

\subsection{Data Fusion}
The estimated results, $\Psi_B$ at the BS and  $\Psi_U$ at the UE, will be fused for estimating the targets' positions and velocities, i.e., $\{\mathbf{q}_l,\mathbf{v}_l, l=1,...,L_t\}$. The data fusion includes two steps:
\begin{itemize}
    \item {\it Target matching:} First, the estimation tuple at the BS $\Psi_B$ needs to be aligned with the estimation tuple $\Psi_U$ at the UE for all the target indices $l=1,...,L_t$. This is achieved by roughly estimating the initial target position sensed by BS, and then calculating the expected latency at the UE. Comparing the expected latency with the estimated latency allows us to match the estimation tuples. 
    \item {\it Joint parameter estimation:} After target matching, the estimated tuple for the $l$th target can be combined and denoted as $\Psi_l=\{\hat{\tau}_{B,l},\hat{f}_{D,l}^{B}, \hat\theta_{B,l},\hat\tau_{U,l}, \hat f_{D,l}^{U} \}$. The joint estimation of $(\mathbf{q}_l, \mathbf{v}_l)$ from $\Psi_l$ needs to  consider the sensing accuracy at the BS and UE, as well as the geometric relationship among BS, UE and a target. The details are elaborated in the following.  
\end{itemize}

\begin{figure}[h]
\vspace{-0.5\baselineskip}
\centerline{\includegraphics[width=0.7\columnwidth]{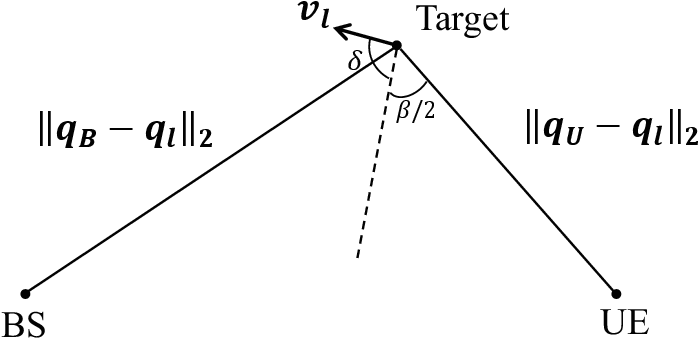}}
\caption{Geometric relationship among BS, UE and a target}
\label{F:Geometry}
\vspace{-0.5\baselineskip}
\end{figure}

For notational convenience, we first convert the estimated delay and Doppler to range and radiate velocity, respectively, which gives $\hat d_l^B=\frac{\hat \tau_{B,l} c_0}{2},\hat d_l^U=\frac{\hat \tau_{U,l} c_0}{2}$ and $\hat v_l^B=\frac{\hat f_{D,l}^Bc_0}{2f_c},\hat v_l^U=\frac{\hat f_{D,l}^Uc_0}{2f_c}$. For angle estimation, in multipath environments, signal coherence can cause rank deficiency in the covariance matrix, preventing the MUSIC algorithm from correctly separating the signal and noise subspaces, thereby introducing angle estimation bias \cite{sturm2011waveform}. Fig.~\ref{F:Geometry} is used to correct imprecise angles. Given $\hat{d}_l^U$ is a numerical value, the positions of the BS and UE are the focus of an ellipse, a circle centered at BS with radius 
$\hat{d}_l^B$ intersects the ellipse at two points. The intersection point, when connected to the BS, forms an angle closer to the estimated angle corresponds to the actual target position. This approach can significantly refine the initial position estimate, even in cases of large angle deviations.

Then, the joint estimation is equivalent to solving the observation model:
\begin{align}
    \bm{y}=\bm{f}(\bm{x};\bm{\varsigma})+\bm{e},\label{eq:Estimodel}
\end{align}
where \(\bm{x} = (\mathbf{q}_{B}, \mathbf{q}_{U})^T\) represents the known quantity, \(\bm{y} = (\hat d_l^B, \hat v_l^B, \hat \theta_{B,l}, \hat d_l^U, \hat v_l^U)^T\) is the observation tuple, \(\bm{\varsigma} = (\mathbf{q}_l, \mathbf{v}_l)^T\) is the set of the unknown parameters to be solved, and the vector \(\bm{e} = (e_{l1}, e_{l2}, e_{l3}, e_{l4}, e_{l5})^T\) includes the errors in each sensing result,  which arise from the OFDM radar resolution and environment influence.

According to the geometric relationship shown in Fig.~\ref{F:Geometry}, the observation model \eqref{eq:Estimodel} can be further expressed as
\begin{align}
\left[\begin{matrix}
\hat d_l^B \\  \hat v_l^B   \\ \hat{\theta}_{B,l} \\ \hat{d}_l^U \\ \hat{v}_l^U
\end{matrix}
\right]=\left[
\begin{matrix}
   \left \| \mathbf{q}_l-\mathbf{q}_{B} \right \|_2 \\
   \frac{\bm{v}_l^T(\mathbf{q}_l-\mathbf{q}_{B})}{\left \| \mathbf{q}_l-\mathbf{q}_{B} \right \|_2}\\
   \arctan\left( \frac{{\mathbf{q}_l(2)-\mathbf{q}_{B}(2)}}{\mathbf{q}_l(1)-\mathbf{q}_{B}(1))} \right)\\
   \left \| \mathbf{q}_l-\mathbf{q}_{B} \right \|_2 + \left \|\mathbf{q}_l-\mathbf{q}_{U} \right \|_2\\
   \left \| \mathbf{v}_l \right \|_2 \cos\delta\cos(\beta/2)
\end{matrix}
\right]+\left[
\begin{matrix}
   e_{l1} \\  e_{l2} \\ e_{l3} \\ e_{l4} \\ e_{l5}
\end{matrix}
\right] \label{eq:ObservationModel}
\end{align}where $e=[e_{l1},e_{l2},e_{l3},e_{l4},e_{l5}]^T$ is the observations' error set.

Then, solving the joint parameter estimation is equivalent to solving the LS problem. 
\begin{equation}\label{eq:LS}
\begin{aligned}
&\arg\min\limits_{q_l, v_l}\quad F= \frac{1}{2} \bm{e}^T \mathbf{W} \bm{e},
\end{aligned}
\end{equation}where $\mathbf{W}\in\mathbb{R}^{5\times 5}$ needs to be adjusted according to the accuracy and importance of the observed values.  In order to solve \eqref{eq:LS}, the initial estimated target positions $\hat{\bm{q}}_{l,0}$ and velocities $\hat{\bm{v}}_{l,0}$ are required. $\hat{\bm{q}}_{l,0}$ can be solved by the BS observed angles and distances by
\begin{equation}
\begin{aligned}
  \hat{\bm{q}}_{l,0} &= \mathbf{q}_B + \hat{d}_l^B
  \begin{bmatrix}
    \cos(\hat\theta_{B,l}) \\
    \sin(\hat\theta_{B,l})
  \end{bmatrix}.
  \label{eq:ql} 
\end{aligned}
\end{equation}

Following that, $\hat{\bm{v}}_{l,0}$ can be solved using equations related to $\hat{v}_l^B$ ,$\hat{v}_l^U$ and $\hat{\bm{q}}_{l,0}$ in \eqref{eq:ObservationModel}. Take $\hat{\bm{q}}_{l,0}$, $\hat{\bm{v}}_{l,0}$ and observations in \eqref{eq:ObservationModel}, the optimized target values are derived via sequential quadratic programming, with the weight matrix $W$ updated using iteratively reweighted least square method (IRLS). The data fusion process for BS-UE cooperative MIMO-OFDM ISAC is summarized in Algorithm 1.
\begin{algorithm}[!h]
    \caption{Data Fusion Algorithm for BS-UE Cooperative MIMO-OFDM ISAC}
    \label{alg:AOS}
    \renewcommand{\algorithmicrequire}{\textbf{Input:}}
    \renewcommand{\algorithmicensure}{\textbf{Output:}}
    
    \begin{algorithmic}[1]
        \REQUIRE Locations of the BS and UE: $\mathbf{q}_B$, $\mathbf{q}_U$;\\
        Observation tuple $\bm y$;\\
        Step tolerance of iteration $\epsilon$; \\
        \STATE Correct potential angle errors.
        \STATE Solve \eqref{eq:ql} to find initial target position \{$\hat{\bm{q}}_{l,0}$\}.
        \STATE Solve initial target velocity \{$\hat{\bm{v}}_{l,0}$\} with $\hat{v}_l^B$ ,$\hat{v}_l^U$ and $\hat{\bm{q}}_{l,0}$.
        \STATE Adjust \{$\hat{\bm{q}_l}$\} and \{$\hat{\bm{v}_l}$\} and update target function as $F'$.
        \IF {$F'-F>\epsilon$}
                \STATE $F$ = $F'$, $\bm{q}'_l=\hat{\bm{q_l}}$, $\bm{v}'_l=\hat{\bm{v}_l}$.
                \STATE Update weight matrix by $W=\frac{1}{\left\vert e \right\vert+10^{-6}}$
                \STATE Return to line 4.
        \ENDIF
        \ENSURE Final optimized values \{$\bm{q}'_l$\}, \{$\bm{v}'_l$\}  
    \end{algorithmic}
\end{algorithm}
 
\begin{table*}[htbp] 
\caption{Coordinates and velocities estimated by the BS and cooperative sensing} \label{resulttable}
\centering
\renewcommand{\arraystretch}{1.5} 
\setlength{\tabcolsep}{6pt}
\begin{tabular}{|c|c|c|c|c|c|c|}
\hline
   & \makecell{Real coordinates\\$\bm{q}_l^T$ (m)} & \makecell{Real velocities\\$\bm{v}_l^T$(m/s)} & \makecell{BS observed coordinates\\$\hat{\bm{q}}_{l,0}^T$(m)}
& \makecell{BS observed velocities\\$\hat{\bm{v}}_{l,0}$(m/s)} & \makecell{\textbf{Jointly estimated}\\\textbf{coordinates $(\bm{q}_l')^T$ (m)}} & \makecell{\textbf{Jointly estimated}\\\textbf{velocities $(\bm{v}_l')^T$ (m/s)}}\\ 
\hline
Target 1 & (59.92,25.06) & (-15,12)& (59.54,24.91)& -9.20&\textbf{(60.18,25.08)} &\textbf{(-14.97,12.03)}\\ \hline
Target 2 & (70.11,14.95) & (20,-10)& (69.21,14.84) & 17.45&\textbf{(69.77,14.92)}&\textbf{(19.98,-10.13)}\\ \hline
Target 3    &(90,30.13) & (0,25) & (90.77,30.55)&7.9&\textbf{(89.75,30.03)}&\textbf{(-0.12,25.0)}\\ 
\hline
\end{tabular} 
\vspace{-0.3cm}
\end{table*}

\section{Numerical Results}
We perform a simulation with carrier frequency $f_c=24$ GHz and a subcarrier spacing of $\Delta f=120$ kHz. The OFDM signal for transmission consists of $N=617$ subcarriers and $M=512$ symbols, with the first 9 and last 8 subcarriers as guard bands, providing sufficient resolution. The receiver samples at rate $B=N\Delta f$. A cyclic prefix of $N_{cp}=149$ is used. The step tolerance $\epsilon=10^{-6}$.
\begin{figure}[b]
\vspace{-0.5\baselineskip}
\centerline{\includegraphics[width=0.8\columnwidth]{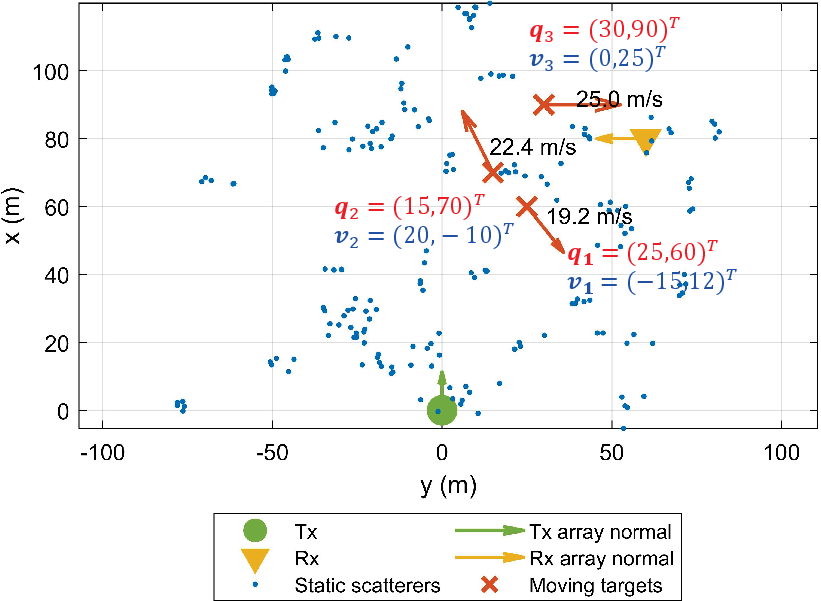}}
\caption{Simulation scenario}
\label{F:Scenario}
\vspace{-\baselineskip}
\end{figure}

Fig.~\ref{F:Scenario} displays the simulation scenario, in which the directions of the transmitter antenna array, receiver antenna array, and targets' information are marked. The BS has 16 antennas with antenna spacing of $\frac{\lambda}{2}$, while the UE has a single antenna. Within the sensing range of the BS and UE, there are three targets and 200 randomly distributed scatterers. The targets have an RCS of 3.5$m^2$ and the scattering coefficients of the scatterers are randomly generated.

After performing BS monostatic and BS-UE bi-static sensing, target matching and joint parameter estimation generate the observation tuples: 
$\bm{y}_1=[64.54,\allowbreak -9.20,\allowbreak 22.60,\allowbreak 105.43,\allowbreak 6.13]$,
$\bm{y}_2=[70.78,\allowbreak 17.45,\allowbreak 12.1,\allowbreak 117.59,\allowbreak -11.46]$,
$\bm{y}_3=[95.77,\allowbreak 7.90,\allowbreak 18.50,\allowbreak 126.08,\allowbreak 7.90]$. Using the proposed data fusion algorithm, the estimated target positions and velocities from BS monostatic and cooperative sensing are shown in Table \ref{resulttable}. From Table \ref{resulttable}, cooperative detection reduces the root mean square error (RMSE) in position estimation by 38\%. Moreover, it enriches the target speed information: by monostatic sensing, we can only get the radial velocity, while using cooperative sensing, the ground-truth velocity can be obtained. 
\begin{figure}
    \vspace{-0.5\baselineskip}
    \centering
    \includegraphics[width=0.95\linewidth]{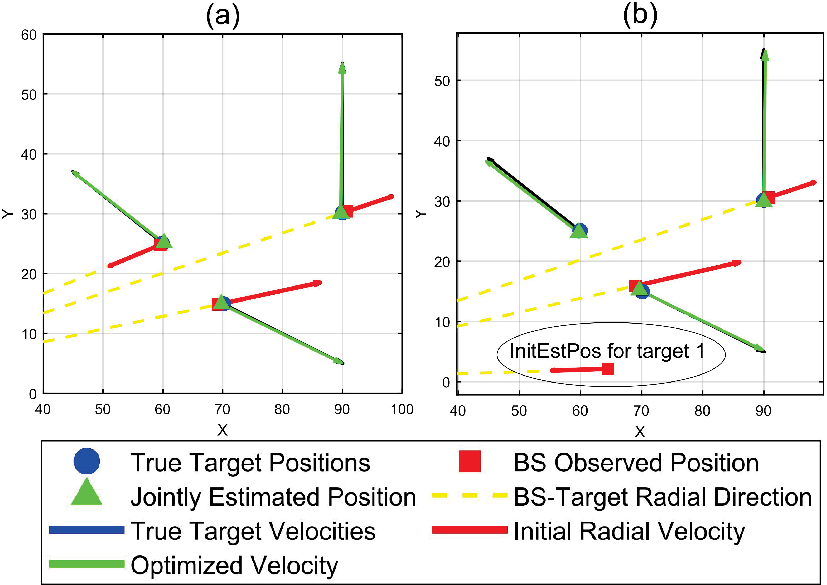}
    \caption{Optimization process of target positions and velocities}
    \label{F:comparison}
    \vspace{-1.5\baselineskip}
\end{figure}

Fig. \ref{F:comparison} shows the optimization process of the proposed data fusion algorithm. In both subfigures, the jointly estimated positions are closer to the true target positions than that observed by the BS. To verify the robustness of the algorithm, the angle estimation situations under good channel conditions and poor channel conditions are discussed respectively. In Fig. \ref{F:comparison}(a), all angles are correctly estimated, which are [$22.6^{\circ}$,$12.1^{\circ}$,$18.5^{\circ}$]. The true values are [$22.69^{\circ}$,$12.09^{\circ}$,$18.43^{\circ}$]. In Fig. \ref{F:comparison}(b), the channel condition is poor and the BS observed position of target 1 deviates greatly from the actual position. The estimated angles are [$1.9^{\circ}$,$12.9^{\circ}$,$18.6^{\circ}$]. Through cooperative sensing, the incorrect estimation resulting from angle error can be rectified. The RMSE reduced 109\%. To ensure statistical validity, we generated 10 additional randomized target configurations under the same conditions, consistently verifying the efficacy of cooperative sensing.

Further, compared with the existing data fusion scheme under the same sensing architecture \cite{Wang2025ISACEC} , our scheme achieves higher velocity estimation accuracy with the same system setting, where the target's distance from the BS is farther than that from the UE while maintaining the target velocity unchanged. The difference lies in that their scheme fuses information from four UEs, and their UEs are the targets.

\section{Conclusions}
This paper proposes a BS-UE cooperative MIMO-OFDM ISAC framework and formulates data fusion as an optimization problem that efficiently solved using the IRLS method. The proposed sensing framework enables more accurate target position estimation than using BS monostatic sensing alone and can solve the target ground-truth velocity, not just the radial velocity. Furthermore, analysis of angle estimation errors and repeated experiments demonstrates that the performance improvement from cooperative sensing becomes more pronounced with poorer initial position estimates, while simultaneously verifying the algorithm's robustness. 

\bibliographystyle{IEEEtran}
\bibliography{conference_10179.bbl}

\begin{thebibliography}{10}
\providecommand{\url}[1]{#1}
\csname url@samestyle\endcsname
\providecommand{\newblock}{\relax}
\providecommand{\bibinfo}[2]{#2}
\providecommand{\BIBentrySTDinterwordspacing}{\spaceskip=0pt\relax}
\providecommand{\BIBentryALTinterwordstretchfactor}{4}
\providecommand{\BIBentryALTinterwordspacing}{\spaceskip=\fontdimen2\font plus
\BIBentryALTinterwordstretchfactor\fontdimen3\font minus
  \fontdimen4\font\relax}
\providecommand{\BIBforeignlanguage}[2]{{%
\expandafter\ifx\csname l@#1\endcsname\relax
\typeout{** WARNING: IEEEtran.bst: No hyphenation pattern has been}%
\typeout{** loaded for the language `#1'. Using the pattern for}%
\typeout{** the default language instead.}%
\else
\language=\csname l@#1\endcsname
\fi
#2}}
\providecommand{\BIBdecl}{\relax}
\BIBdecl

\bibitem{ITU}
ITU-R, ``Framework and overall objectives of the future development of {IMT}
  for 2030 and beyond,'' [Online],
  \url{https://www.itu.int/en/ITU-R/study-groups/rsg5/rwp5d/imt2030/Pages/default.aspx/}.

\bibitem{3GPP_ISAC}
{3GPP TR 22.837}, ``Feasibility study on integrated sensing and
  communication,'' {3rd Generation Partnership Project (3GPP)}, Tech. Rep.,
  2022.

\bibitem{Brunner2024BistaticOI}
D.~Brunner, L.~G. de~Oliveira, and C.~M. et~al., ``Bistatic ofdm-based isac
  with over-the-air synchronization: System concept and performance analysis,''
  \emph{IEEE Trans. Microw. Theory Tech.}, vol.~1, no.~1, pp. 1--1, 2024.

\bibitem{Liu2021IntegratedSA}
F.~Liu, Y.~Cui, C.~Masouros, J.~Xu, T.~X. Han, Y.~C. Eldar, and S.~Buzzi,
  ``Integrated sensing and communications: Toward dual-functional wireless
  networks for 6g and beyond,'' \emph{IEEE Journal on Selected Areas in
  Communications}, vol.~40, pp. 1728--1767, 2021.

\bibitem{Xie2022CollaborativeSI}
L.~Xie, S.~Song, Y.~C. Eldar, and K.~B. Letaief, ``Collaborative sensing in
  perceptive mobile networks: Opportunities and challenges,'' \emph{IEEE
  Wireless Commun.}, vol.~30, no.~1, pp. 16--23, 2022.

\bibitem{Jiang2023CooperationBasedJA}
W.~Jiang, Z.~Wei, S.~Yang, Z.~Feng, and P.~Zhang, ``Cooperation-based joint
  active and passive sensing with asynchronous transceivers for perceptive
  mobile networks,'' \emph{IEEE Trans. Wireless Commun.}, vol.~23, no.~10, pp.
  15\,627--15\,641, 2023.

\bibitem{Zhou2024JointTD}
L.~Zhou, J.~Dai, W.~Xu, and C.~Chang, ``Joint target detection and channel
  estimation for distributed massive mimo isac systems,'' \emph{IEEE Trans.
  Cogn. Commun. Netw.}, vol.~11, no.~1, pp. 300--315, 2025.

\bibitem{9727176}
J.~Yang, C.-K. Wen, and S.~Jin, ``Hybrid active and passive sensing for slam in
  wireless communication systems,'' \emph{IEEE J. Sel. Areas Commun.}, vol.~40,
  no.~7, pp. 2146--2163, 2022.

\bibitem{10570616}
Y.~Ji and X.~Xu, ``Age-optimal joint sampling and transmitting scheduling for
  wireless sensor networks with energy harvesting,'' in \emph{Proc. IEEE
  Wireless Commun. Netw. Conf. (WCNC)}, 2024, pp. 1--6.

\bibitem{b5}
Z.~Peng, C.~Li, and F.~Uysal, \emph{Modern Radar for Automotive
  Applications}.\hskip 1em plus 0.5em minus 0.4em\relax Radar Sensor Technology
  XXVII, 2022.

\bibitem{sturm2011waveform}
C.~Sturm and W.~Wiesbeck, ``Waveform design and signal processing aspects for
  fusion of wireless communications and radar sensing,'' \emph{Proceedings of
  the IEEE}, vol.~99, no.~7, pp. 1236--1259, 2011.

\bibitem{Wang2025ISACEC}
Y.~Wang, K.~Zu, L.~Xiang, Q.~Zhang, Z.~Feng, J.~Hu, and K.~Yang, ``Isac enabled
  cooperative detection for cellular-connected uav network,'' \emph{IEEE Trans.
  Wireless Commun.}, vol.~24, pp. 1541--1554, 2025.

\end{thebibliography}
\end{document}